\documentclass[a4paper,11pt]{article}
\usepackage{pos}
\usepackage{dsfont}

\newcommand{\ket}[1]{\mbox{$ | #1 \rangle $}}
\newcommand{\bra}[1]{\mbox{$ \langle #1 | $}}
\newcommand{\OO}{\ensuremath{\mathcal{O}}}
\newcommand{\Id}{\ensuremath{\mathds{I}}}

\newcommand{\tr}{\ensuremath{\textrm{Tr}}}

\newcommand{\UU}{\mathcal{U}}

\graphicspath{{figures/}}

\begin{document}

\title{Model-Independent Error Mitigation in Parametric Quantum Circuits and Depolarizing Projection of Quantum Noise}
\ShortTitle{Model-Independent Error Mitigation and Depolarizing Projection of Quantum Noise}

\author*[a]{Xiaoyang Wang}
\author[a]{Xu Feng}
\affiliation[a]{School of Physics and Center of High Energy Physics, Peking University, Beijing 100871, China}

\author[b,c]{Lena Funcke}
\affiliation[b]{Center for Theoretical Physics, Co-Design Center for Quantum Advantage, and NSF AI Institute for Artificial Intelligence and Fundamental Interactions, Massachusetts Institute of Technology, 77 Massachusetts Avenue, Cambridge, MA 02139, USA}
\affiliation[c]{Perimeter Institute for Theoretical Physics, 31 Caroline Street North, Waterloo, ON N2L 2Y5, Canada}

\author[d]{Tobias Hartung}
\affiliation[d]{Department of Mathematics, King’s College London, Strand, London WC2R 2LS, United Kingdom}

\author[e]{Karl Jansen}
\affiliation[e]{Deutsches Elektronen-Synchrotron DESY, Platanenallee 6, 15738 Zeuthen, Germany}

\author[f]{Stefan K{\"u}hn}
\affiliation[f]{Computation-Based  Science  and  Technology  Research  Center, The  Cyprus  Institute,  20  Kavafi  Street,  2121  Nicosia,  Cyprus}

\author[f,g]{Georgios Polykratis}

\affiliation[g]{Department of Physics, University of Cyprus, P.O. Box 20537, 1678 Nicosia, Cyprus}

\author[e,h]{Paolo Stornati}
\affiliation[h]{ICFO, The Barcelona Institute of Science and Technology, Av. Carl Friedrich Gauss 3, 08860 Castelldefels (Barcelona), Spain
}

\emailAdd{zzwxy@pku.edu.cn}
\emailAdd{xu.feng@pku.edu.cn}
\emailAdd{lfuncke@mit.edu}
\emailAdd{tobias.hartung@desy.de}
\emailAdd{karl.jansen@desy.de}
\emailAdd{s.kuehn@cyi.ac.cy}
\emailAdd{g.polykratis@cyi.ac.cy}
\emailAdd{paolo.stornati@icfo.eu}

\abstract{
Finding ground states and low-lying excitations of a given Hamiltonian is one of the most important 
problems in many fields of physics. As a novel approach, quantum computing on Noisy Intermediate-Scale 
Quantum (NISQ) devices offers the prospect to efficiently perform such computations and may eventually outperform 
classical computers. However, current quantum devices still suffer from inherent quantum noise. In this work, we propose an error mitigation scheme suitable for 
parametric quantum circuits. This scheme is based on projecting a general quantum noise channel onto 
depolarization errors. Our method can efficiently reduce errors in quantum computations, which we demonstrate by carrying out simulations both on classical and IBM's quantum devices. In particular, we test the performance of the method by computing the mass gap of the transverse-field Ising model using the variational quantum eigensolver algorithm.\\\\ Preprint number: MIT-CTP/5354 
}

\FullConference{%
 The 38th International Symposium on Lattice Field Theory, LATTICE2021
  26th-30th July, 2021
  Zoom/Gather@Massachusetts Institute of Technology
}

\maketitle

\section{Introduction}

Recent advances in quantum technologies open up a new route to tackle quantum many-body problems. In particular, simulations on quantum devices are based on the Hamiltonian formulation and hence have the potential to circumvent the sign problem and allow for simulating real-time dynamics. Thus, they offer the prospect to study problems that are very hard or even inaccessible for the conventional Monte Carlo approach. First successful proof-of-principle experiments~(see, e.g., \cite{Martinez2016,Kokail2018,Klco2018,Klco2019,Ciavarella2021,Zhou:2021kdl}) render this approach particularly promising for the future. 

Current noisy intermediate-scale quantum (NISQ) devices~\cite{Preskill:2018} are still suffering from a considerable amount of quantum noise, which limits the depth of the quantum circuits that can be executed faithfully. While full quantum error correction is presently still under development, errors can be partially mitigated using error mitigation schemes~(see, e.g., \cite{Kandala2017,Endo2018,funcke2020measurement,Geller2021,Alexandrou:2021ynh,Alexandrou:2021wqu}). Here, we propose a novel scheme to mitigate quantum noise, which is inspired by randomized benchmarking (RB)~\cite{PhysRevA.85.042311} and randomized compiling (RC)~\cite{PhysRevA.94.052325}. RB is used to characterize the error rates of quantum channels, while RC projects arbitrary quantum noise into Pauli channels. In our study, we extend the scope of RC to further project Pauli channels into depolarizing channels, by utilizing the mathematical formalism used for RB. This depolarizing projection allows us to perform error mitigation using a recently developed technique for mitigating depolarizing errors~\cite{vovrosh2021efficient}. We test the efficiency of the proposed depolarizing projection and prove that the required depth of the quantum circuit grows polynomially with the number of qubits. We also implement our mitigation technique to calculate the spectrum of the transverse-field Ising Hamiltonian using both classical and quantum devices. We find that, within statistical errors, both the classical and quantum results converge to the ones obtained from exact diagonalization of the Hamiltonian.

\section{Mitigation of depolarizing noise}\label{sec:depolarizing}

In this section, we will provide an introduction to quantum noise and review a recently developed mitigation technique for depolarizing errors~\cite{vovrosh2021efficient}. Suppose we have an $N$-qubit quantum system living in the Hilbert space $\mathcal{H}$ with dimension $d=2^N$. We define the set of density operators $D(\mathcal{H})$ as non-negative, unit-trace, linear operators on $\mathcal{H}$. With a given density operator $\rho\in D(\mathcal{H})$, the expectation value of an observable $\mathcal{O}$ is defined as $\langle\mathcal{O}\rangle\equiv\tr(\mathcal{O}\rho)$. For the example of calculating an $n$-th eigenvalue $E_n$ of a Hamiltonian $H$, the density operator is given by $\rho = \ket{E_n}\bra{E_n}$, such that
\begin{align}
    \langle H\rangle_n=\tr(H\rho)=\bra{E_n}H\ket{E_n}=E_n.
\end{align}
If one can prepare the exact density operator $\ket{E_n}\bra{E_n}$ using the variational quantum eigensolver (VQE) algorithm~\cite{Peruzzo2014}, $E_n$ can be correctly estimated. However, quantum noise will deform $\rho$ into a noisy density operator $\rho'$, such that the noise-contaminated expectation value $\langle H\rangle'\equiv\tr(H\rho')$ will be different from $E_n$. To mitigate this problem, we will in the following discuss the origin of the quantum noise that yields $\rho'$, starting from a noiseless quantum circuit to prepare $\rho$.

In a quantum computer, one uses a quantum circuit to prepare the final density operator $\rho$ from an initial density operator $\rho_0$, where $\rho_0$ is usually taken to be $\ket{0}\bra{0}^{\otimes N}$ with $N$ denoting the number of qubits. Subsequently, one measures $\rho$ in the computational basis $\{\ket{0},\ket{1}\}^{\otimes N}$. We divide the state preparation circuit into different circuit layers, where each layer consists of a set of simultaneously implemented logical gates. In the ansatz circuit that we use for the VQE, a circuit layer can either be a layer of single-qubit rotation gates or a layer of (non-overlapping) two-qubit entangling CNOT gates. Thus, the circuit for preparing a noiseless state can be written as
\begin{align}
    \UU_L = u_{L-1}\ldots u_0 = \prod_{i=0}^{L-1}u_i,
    \label{eq:circuit-layers}
\end{align}
where $u_i$ denotes the different circuit layers and $L$ is the total number of circuit layers. The final density operator $\rho$ can be obtained by 
\begin{align}
    \rho=\hat{\UU}_L(\rho_0)=\prod_{i=0}^{L-1}\hat{u}_i(\rho_0)=\prod_{i=0}^{L-1}u_i \rho_0 u^{\dagger}_i,
\end{align}
where $\hat{\UU}_L=\prod_{i=0}^{L-1}\hat{u}_i$ denotes the super-operator and $\hat{u}_i$ transforms $\rho_0$ as $\hat{u}_i(\rho_0)= u_i \rho_0 u^{\dagger}_i$.

Next, we go beyond the noise-free case and consider quantum noise in the state preparation circuit. Quantum noise linearly maps the set $D(\mathcal{H})$ of density operators to itself, while keeping the trace of the density operators equal to 1. These maps $\mathcal{E}$ are called quantum channels, which are completely positive, trace-preserving (CPTP) maps and can be expanded in terms of the Kraus decomposition~\cite{Nielsen2000}
\begin{align}
  \mathcal{E}(\rho)=\sum_{k=1}^K A_k \rho A_k^{\dagger},\quad \forall \rho\in D(\mathcal{H}).
\end{align}
 Here, $A_k$ are Kraus operators of dimension $d\times d$, and the trace of $\mathcal{E}(\rho)$ is preserved due to the completeness relation $\sum_{k=1}^K A_k^{\dagger} A_k=\Id$. It can be shown that one can always find a decomposition with $K=d^2$, such that the decomposition contains $d^4-d^2$ degrees of freedom~\cite{Nielsen2000}. For example, for $N=2$ qubits, there are $d^4-d^2=2^{4N}-2^{2N}=240$ free parameters to be determined. 

Quantum noise deforms the noiseless circuit layers $\hat{u}_i$ into noisy circuit layers $\hat{u}_i'$. Here and in the following, we denote noisy quantities with a prime symbol $(')$. The noisy layers $\hat{u}_i'$ can be decomposed into
\begin{align}
   \hat{u}'_i(\rho)= \mathcal{E}_i \circ \hat{u}_i(\rho)=\sum_{k=1}^K A_{i,k} u_i \rho u_i^{\dagger} A_{i,k}^{\dagger}.
    \label{eq:noisy-layers}
\end{align}
Thus, the final density operator will be deformed by quantum noise into 
\begin{align}
    \rho'=\hat{\UU}_L'(\rho_0)=\prod_{i=0}^{L-1}\hat{u}_i'(\rho_0).
\end{align}
The resulting noisy expectation value $\tr(\mathcal{O}\rho')$ hence deviates from the noiseless one $\tr(\mathcal{O}\rho)$. This deviation is generally determined by all the degrees of freedom of all quantum channels $\mathcal{E}_i$, unless it is possible to approximate the resulting effect with a simpler channel, as we will demonstrate in the next section.

To mitigate the effects of quantum noise, one needs to relate the noisy expectation values to their noiseless counterparts. While this is challenging for generic quantum channels, some specific quantum channels can be handled more easily. One example is the depolarizing channel
\begin{align}
    \rho'= \mathcal{E}^D_r(\rho)=(1-r)\rho+r\frac{\Id}{d},
    \label{eq:depolarizing-channel}
\end{align}
which transforms the density operator $\rho$ into the maximally mixed state  $\Id/d$ with some depolarizing probability $r$. Assuming that all quantum channels $\mathcal{E}_i$ in Eq.~\eqref{eq:noisy-layers} are depolarizing~\cite{vovrosh2021efficient}, the final density operator will be of the form given by Eq.~\eqref{eq:depolarizing-channel}, with an unknown depolarizing probability $r$. The noiseless expectation value of the observable $\mathcal{O}$ can then be obtained from the noisy one using
\begin{align}
    \tr(\OO \rho)=\frac{\tr(\OO \rho')}{1-r},
    \label{eq:mitigation-formula}
\end{align}
if we assume that $\mathcal{O}$ is traceless, which for instance is the case for the Hamiltonian of the transverse-field Ising model (see also Sec.~\ref{sec:experimental}). The only unknown parameter $r$ in Eq.~\eqref{eq:mitigation-formula} can be obtained by measuring purities $\tr(\rho'^2)$ using state tomography~\cite{Nielsen2000} or randomized measurements~\cite{PhysRevLett.120.050406}.

Note that the assumption of having a depolarizing channel after each circuit layer is generally not fulfilled for real quantum devices. However, we will prove in the next section that for specific quantum circuits, more general quantum channels can be projected into depolarizing ones.

\section{Depolarizing projection of quantum channels}\label{sec:projection}

In this section, we aim to demonstrate that specific quantum channels can be projected into depolarizing ones. For simplicity, let us consider an incoherent Pauli channel
\begin{align}
  \mathcal{E}_p(\rho)=\sum_{a\in \mathbf{P}^N} p_a \hat{P}_a(\rho)=(1-p)\Id+p\Lambda,
  \label{eq:pauli-channel}
\end{align}
where $\mathbf{P}^N=\{I,X,Y,Z\}^{\otimes N}$ is the set of $N$-qubit Pauli matrices, $\hat{P}_a$ is the super-operator of an element in $\mathbf{P}^N$, $p_a$ is the Pauli error probability, $p\equiv \sum_{a\in \mathbf{P}^N/\Id}p_a$ is the total error probability of the non-trivial Pauli operators, $\Id$ is the identity that acts trivially on density matrices, and $\Lambda$ is the full Pauli channel except $\Id$. Note that the incoherent Pauli channel in Eq.~\eqref{eq:pauli-channel} is a more general quantum channel than the depolarizing channel in Eq.~\eqref{eq:depolarizing-channel}, because the depolarizing channel can be derived by setting all Pauli error probabilities $p_a$ equal except for the one multiplying the identity. 

In the following, we will model arbitrary quantum noise by the Pauli channel in Eq.~\eqref{eq:pauli-channel}. This simplification can be justified by randomized compiling~\cite{PhysRevA.94.052325}, which projects arbitrary quantum noise into its corresponding Pauli channel by dressing gates with a twirling set. Recent experiments verified that this projection works extremely well~\cite{Ware_2021}. In addition, we assume that the quantum noise is layer-independent, time-stationary, and Markovian (LTM), similar to the gate-independent, time-stationary and Markovian noise introduced in~\cite{Flammia_2020}. 

Regarding the definition of LTM noise, a noisy implementation of a circuit layer $\hat{u}'(t)$ is \textit{time-stationary} if the linear noise map is independent of $t$. To make this noisy implementation \textit{layer-independent} and \textit{Markovian}, we can construct $\hat{u}'=\mathcal{E}\circ \hat{u}$ such that the CPTP map $\mathcal{E}$ has no dependence on $\hat{u}$ (layer-independent) or other parts of the quantum circuit (Markovian). 

Now, let us consider a quantum circuit for preparing a state that is contaminated by LTM noise, similar to the noiseless circuit in Eq.~\eqref{eq:circuit-layers}. Such a circuit can be represented by a linear map
\begin{align}
    \hat{\UU}'_L= \prod_{i=0}^{L-1}\mathcal{E}_p \circ \hat{u}_i,
\end{align}
where $\mathcal{E}_p$ is defined in Eq.~\eqref{eq:pauli-channel}. Substituting this definition for $\mathcal{E}_p$, the linear map can be expanded similar to the binomial expansion. We find
\begin{align}
    \hat{\UU}'_L= P(0)\hat{\UU}_L+\sum_{M=1}^L P(M) \hat{\UU}_L\frac{1}{{L \choose M}}\prod_{\alpha=0}^{M-1} \left(\sum_{i_{\alpha}=0}^{i_{\alpha+1}-1} \mathcal{D}_{i_\alpha}^{\dagger}\; \Lambda\; \mathcal{D}_{i_\alpha}\right),
    \label{eq:full-noisy-unitary}
\end{align}
where $\mathcal{D}_{i_{\alpha}}\equiv \hat{u}_{i_\alpha}\dots\hat{u}_{0}$ is the product of the circuit layers, and we defined $L\equiv i_M$. The error probabilities $P(M)$ follow the binomial distribution
\begin{align}
    P( M ) = {L \choose M}p^M(1-p)^{L-M},
    \label{eq:binomial-distribution}
\end{align}
such that the noiseless term in Eq.~\eqref{eq:full-noisy-unitary} is given by $P(0)=P(M=0)$. Note that each product $\prod_{\alpha=0}^{M-1} (\sum_{i_{\alpha}=0}^{i_{\alpha+1}-1}\mathcal{D}_{i_\alpha}^{\dagger}\; \Lambda\; \mathcal{D}_{i_\alpha})$ with $M\in 1,2,\ldots L$ contains ${L \choose M}$ terms; therefore, the expression in Eq.~\eqref{eq:full-noisy-unitary} is normalized by a factor of ${L \choose M}$.

Next, we assume that the product of the total Pauli error probability $p$ and the number of the circuit layers $L$ is small, $pL\ll 1$. In this case, we can approximate Eq.~\eqref{eq:binomial-distribution} by noting that $P(M=1)\sim pL \gg P(M>1)\sim O(p^2L^2)$. Using this approximation, we only need to consider the lowest-order term in Eq.~\eqref{eq:full-noisy-unitary} with $M=1$, which yields
\begin{align}
    \hat{\UU}'_L=(1-pL)\hat{\UU}_L+pL\hat{\UU}_L\frac{1}{L}\sum_{i_0=0}^{L-1} \mathcal{D}_{i_0}^{\dagger}\; \Lambda \; \mathcal{D}_{i_0}+O(p^2L^2).
    \label{eq:1}
\end{align}
Finally, we assume that the set of circuit layer products $\{\mathcal{D}_{i_0}\}$ with $i_0=0,\ldots L-1$ is random and that $L$ is large enough so that $\{\mathcal{D}_{i_0}\}$ is a unitary 2-design~\cite{PhysRevA.80.012304}. Thus, the second term in Eq.~\eqref{eq:1} reduces to
\begin{align}
  \frac{1}{L}\sum_{i_0=0}^{L-1} \mathcal{D}_{i_0}^{\dagger}\; \Lambda\; \mathcal{D}_{i_0}(\rho_0)  = q\rho_0 +(1-q)\frac{\Id}{d}\equiv \rho_d,
    \label{eq:U-2-design}
\end{align}
where $q=-(d^2-1)^{-1}$ depends on the dimension $d=2^N$ of the Hilbert space. Note that the typical cardinality of a unitary 2-design, such as the Clifford group, is exponentially large with respect to the number $N$ of qubits. Thus, the required number of circuit layers may limit the efficiency of this twirling effect. However, we will prove in the next section that the number of circuit layers required to estimate the expectation value of some observable $\mathcal{O}$ to a fixed precision has no explicit dependence on the number of qubits. In contrast, the required number of circuit layers will only be proportional to the norm squared of the observable under consideration.

Using Eqs.~\eqref{eq:1} and \eqref{eq:U-2-design}, we obtain our final expression for the noise-contaminated prepared state,
\begin{align}
    \rho'=\hat{\mathcal{U}}'_L(\rho_0)=[1-pL(1-q)]\rho+pL(1-q)\frac{\Id}{d}+O(p^2L^2),
    \label{eq:final-noisy-result}
\end{align}
where $\rho=\hat{\UU}_L(\rho_0)$ is the noiseless pure state. Thus, we have projected the Pauli channel in Eq.~\eqref{eq:pauli-channel} into the depolarizing channel in Eq.~\eqref{eq:depolarizing-channel}, up to errors of order $O(p^2L^2)$. This implies that we can mitigate arbitrary quantum noise and estimate the true expectation value of observables with the error mitigation technique discussed in the previous section, in particular Eq.~\eqref{eq:mitigation-formula}.

\section{Numerical verification and error estimation}\label{sec:verification}

In our derivation in the previous section, the key assumption is the unitary 2-design that yields Eq.~\eqref{eq:U-2-design}. In the following, we will numerically verify this equation and investigate the rate of convergence to the unitary 2-design depending on the number $L$ of circuit layers. In particular, we will examine the generated density matrix, assuming LTM quantum noise. 

Let us consider the density matrix $\rho_d$ given by Eq.~\eqref{eq:U-2-design} and compute its von-Neumann entropy
\begin{align}
  S(\rho_d)\equiv -\tr(\rho_d \log \rho_d)&=\log(d+1)+\frac{d}{d+1}\log\left(\frac{d-1}{d}\right).
  \label{eq:theoretical-entropy}
\end{align}
This theoretical expectation needs to be compared with numerical results, which can be obtained using the following steps. First, we prepare a sequence of unitary matrices sampled from the circular unitary ensemble. Second, we insert one Pauli operator behind one randomly chosen unitary matrix and apply this combined operator to the initial state $\ket{0}^{\otimes N}$. Third, we calculate the resulting density matrix $\rho$. Fourth, we repeat the second and third step $T$ times and calculate the entropy of the averaged density matrix $S(\sum \rho/T)$ with exact diagonalization. 

\begin{figure*}[b]
  \centering
  \includegraphics[width=\columnwidth]{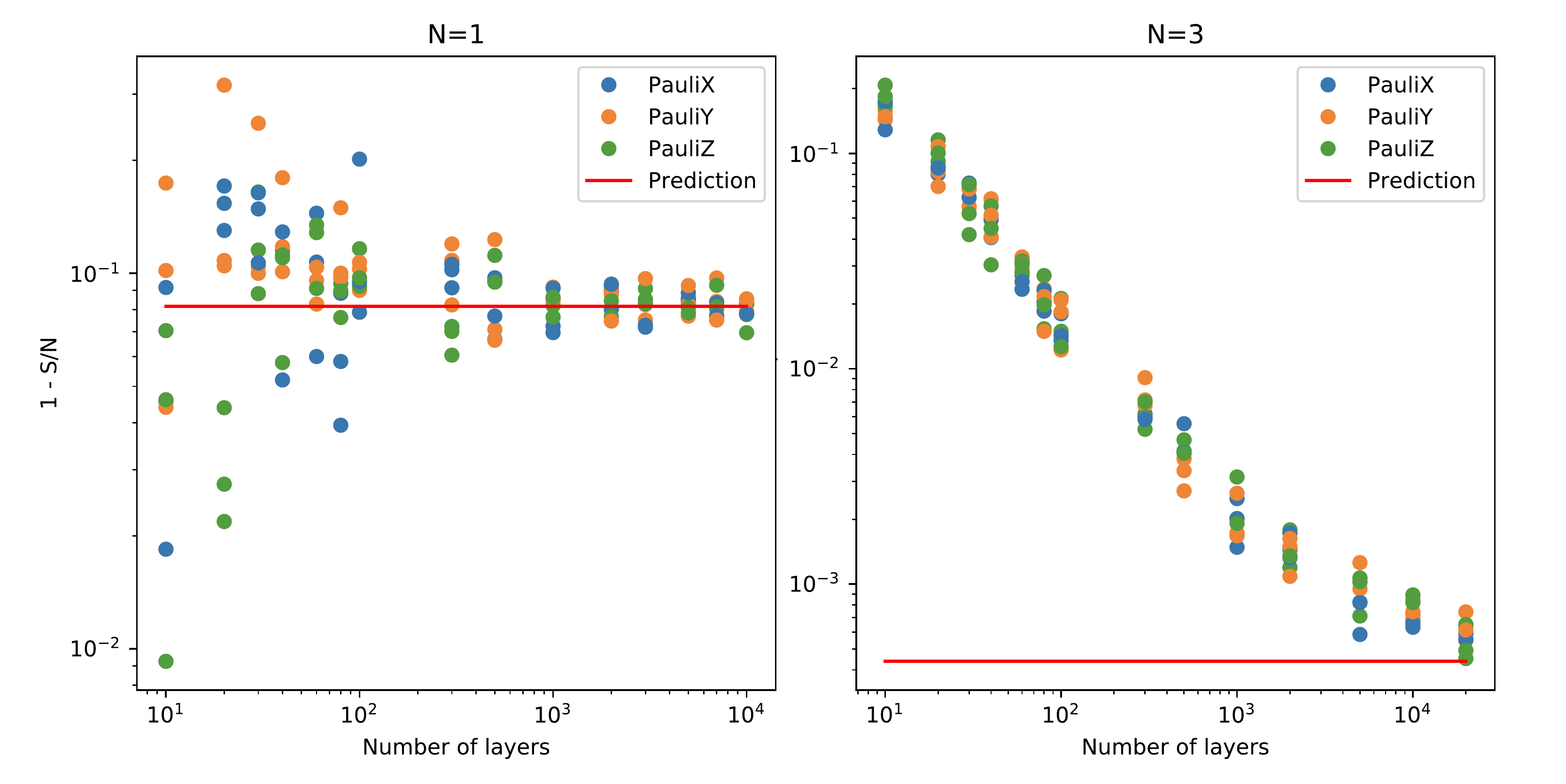}
  \caption{Numerical results for $1-S/N$, where $S$ is the entropy and $N$ is the number of qubits, as a function of the number of layers $L$. We plot results for three Pauli channels, with $\Lambda=X$ (blue points), $\Lambda=Y$ (orange points), $\Lambda=Z$ (green points), as well as the theoretical prediction (red line) from Eq.~\eqref{eq:theoretical-entropy} for $N=1$ (left panel) and $N=3$ (right panel) qubits. The entropies converge to the predicted value with increasing $L$. For each data point, we generate a sequence of random unitaries according to the Haar measure and then randomly apply a single-qubit Pauli operator ($X$, $Y$, or $Z$) to the first qubit after one of the unitaries.}
  \label{fig:single-insertion} 
\end{figure*}

In Fig.~\ref{fig:single-insertion}, we show the numerical results for the entropy as a function of the number of layers, where we used $N=1$ qubit (left panel) and $N=3$ qubits (right panel). We plot results for three different Pauli channels, with $\Lambda=X$ (blue points), $\Lambda=Y$ (orange points), and $\Lambda=Z$ (green points). For all of these channels, we observe that our results for $1-S/N$, which measures the distance of the averaged density matrix to the maximally mixed state, converges to the theoretical value (red line) as predicted in Eq.~\eqref{eq:theoretical-entropy}. 

From our numerical experiments, we see that the number of layers required for complete convergence to the unitary 2-design can in principle be very large. This is expected because the cardinality of a typical unitary 2-design, such as the Clifford group, on $N$-qubit systems scales as $O(4^{N^2})$. However, similar to the scalability of the RB protocol~\cite{PhysRevA.85.042311}, this will not be a practical obstacle if our aim is to estimate the expectation value of an observable and mitigate the error according to the depolarizing projection. 
Suppose our desired confidence level is $1-\delta$, where $\delta$ is some constant, e.g., $\delta=0.05$. Following~\cite{PhysRevA.85.042311}, the number of circuit layers required to reach this confidence level is given by
\begin{align}
    L=\frac{\ln(\frac{2}{\delta})||H||_{\infty}^2}{2\epsilon^2},
    \label{eq:L}
\end{align}
where $\epsilon$ denotes the accuracy of the estimate and $||H||_{\infty}$ is the operator-norm (Schatten-$\infty$ norm) of a given Hamiltonian. $||H||_{\infty}$ is typically $O({\rm poly}(N))$ for most quantum many-body problems, which proves the scalability of controlling the error in the Hamiltonian expectation value estimation.

\section{Experimental results with VQE}\label{sec:experimental}

\begin{figure}[t]
\centering
  \includegraphics[width=0.85\columnwidth]{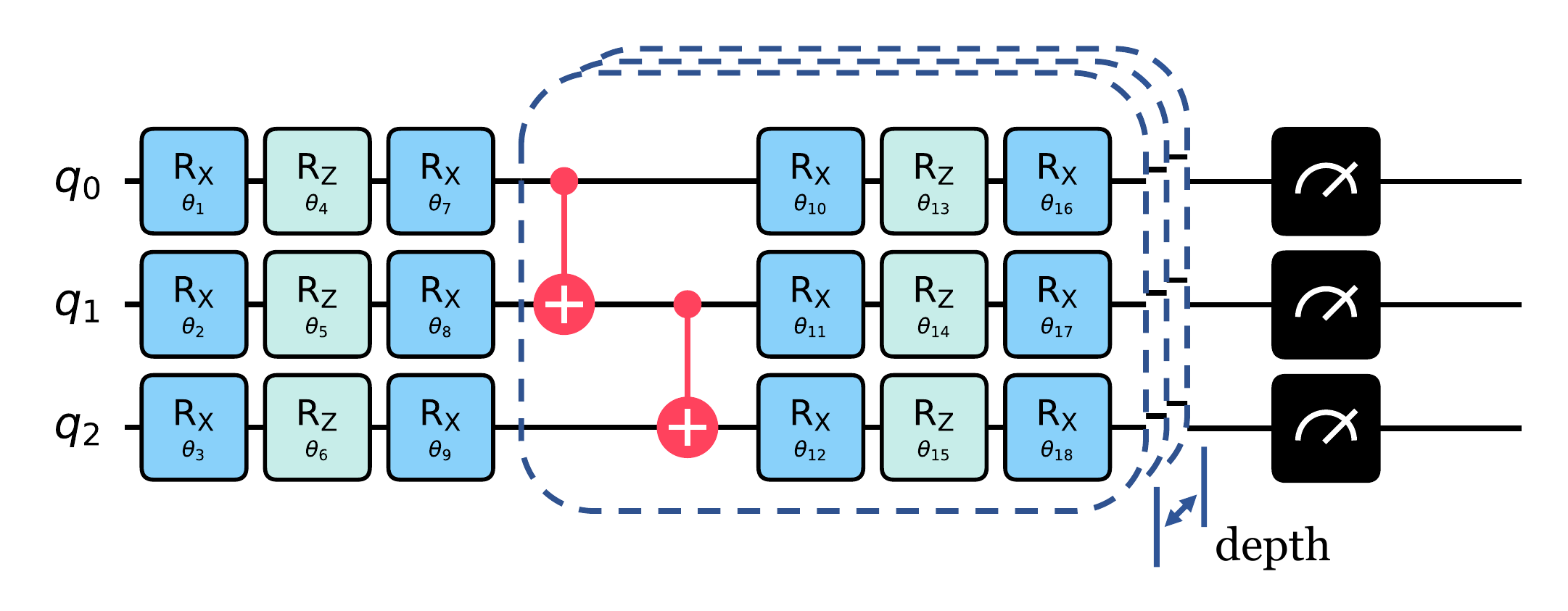}
  \caption{VQE circuit ansatz used in our experiments. The blue and green $R_X$ and $R_Z$ boxes denote parametric rotation gates, the red two-qubit connections are CNOT gates, the black boxes are the final measurements, and the dashed blue lines correspond to different layers of the quantum circuit.}
  \label{fig:EfficientSU2}
\end{figure}

The VQE algorithm~\cite{Peruzzo2014} is a hybrid quantum-classical algorithm for finding the low-lying eigenstates of a Hamiltonian using a variational approach, where the computing-intensive cost function is evaluated on the quantum device. The VQE approach is an ideal platform to test our mitigation scheme, as the circuit layers of the VQE ansatz are sufficiently random to enable the projection of quantum noise to the depolarizing channel (see Sec.~\ref{sec:projection}). In the following, we use the VQE algorithm to estimate the ground-state energy of the transverse-field Ising model
\begin{align}
  H = x \sum_{i=0}^{N-1} X_i X_{i+1} -\sum_{i=0}^{N-1} Z_i.
  \label{eq:transversal_ising_hamiltonian}
\end{align}
Here, $x$ is the coupling strength between nearest-neighbour spinors, and we use periodic boundary conditions, i.e., $X_N=X_0$. 
To approximate the ground state $\ket{\psi}$ of the Hamiltonian in Eq.~\eqref{eq:transversal_ising_hamiltonian}, we use the parametric VQE ansatz $\ket{\psi(\mbox{\boldmath$\theta$})}$ prepared with the quantum circuit depicted in Fig.~\ref{fig:EfficientSU2}. The rotation parameters $\mbox{\boldmath$\theta$} = (\theta_1, \theta_2, \dots)$ are optimized using a variational approach to make the expectation value $E(\mbox{\boldmath$\theta$})\equiv\bra{\psi(\mbox{\boldmath$\theta$})}H\ket{\psi(\mbox{\boldmath$\theta$})}$ as small as possible.  

Even for a sufficiently expressive ansatz circuit, the VQE approach has two main sources of error. On the one hand, the minimization of the variational parameters is not guaranteed to converge to the global minimum  $\mbox{\boldmath$\theta$}_0$ of the cost function. Thus, the resulting state $\ket{\psi(\mbox{\boldmath$\theta$})}$ might not correspond to the exact ground state of the Hamiltonian. On the other hand, even if the global minimum is found during the optimization process, the quantum noise during the execution of the circuit will in general yield an energy expectation value $E_0 =\bra{\psi(\mbox{\boldmath$\theta$}_0)}H\ket{\psi(\mbox{\boldmath$\theta$}_0)}$ that deviates from the exact ground state energy. Since our mitigation scheme is concerned with the errors caused by the inherent noise of the quantum device, we ensure that the parameter set we obtain is sufficiently close to the global minimum. To this end, we also compute the exact state vector $\ket{\psi(\mbox{\boldmath$\theta$})}$ for each parameter set along the minimization procedure, and check the overlap of with the true ground state computed with exact diagonalization, $|\langle \psi(\mbox{\boldmath$\theta$})\ket{\psi}|$. We run the optimization until the resulting overlap is larger than $0.99$, such that the systematic error due to the final parameters deviating from the global minimum is negligible.

\begin{figure}[htp!]
\centering
  \includegraphics[width=0.49\columnwidth]{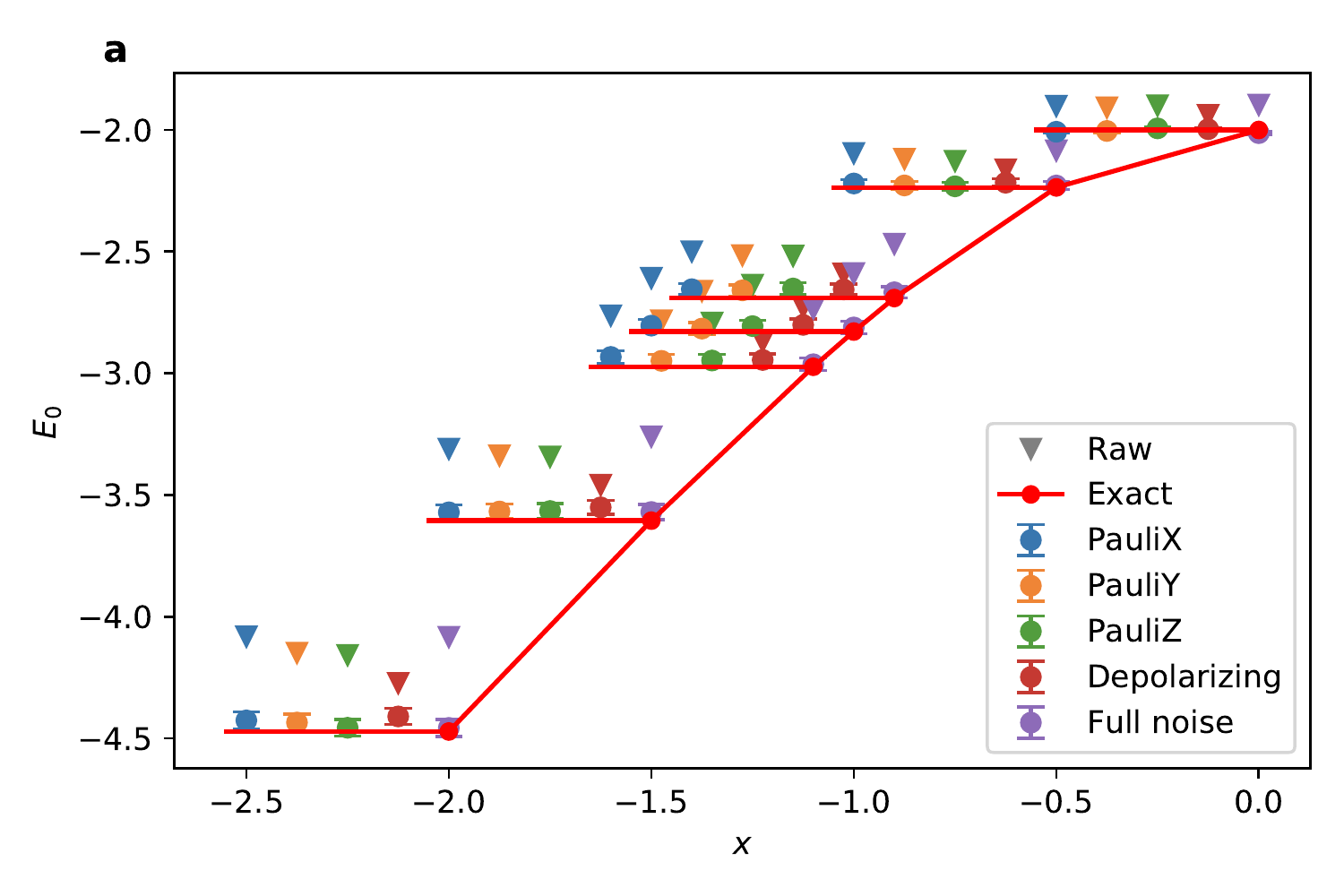}
    \includegraphics[width=0.49\columnwidth]{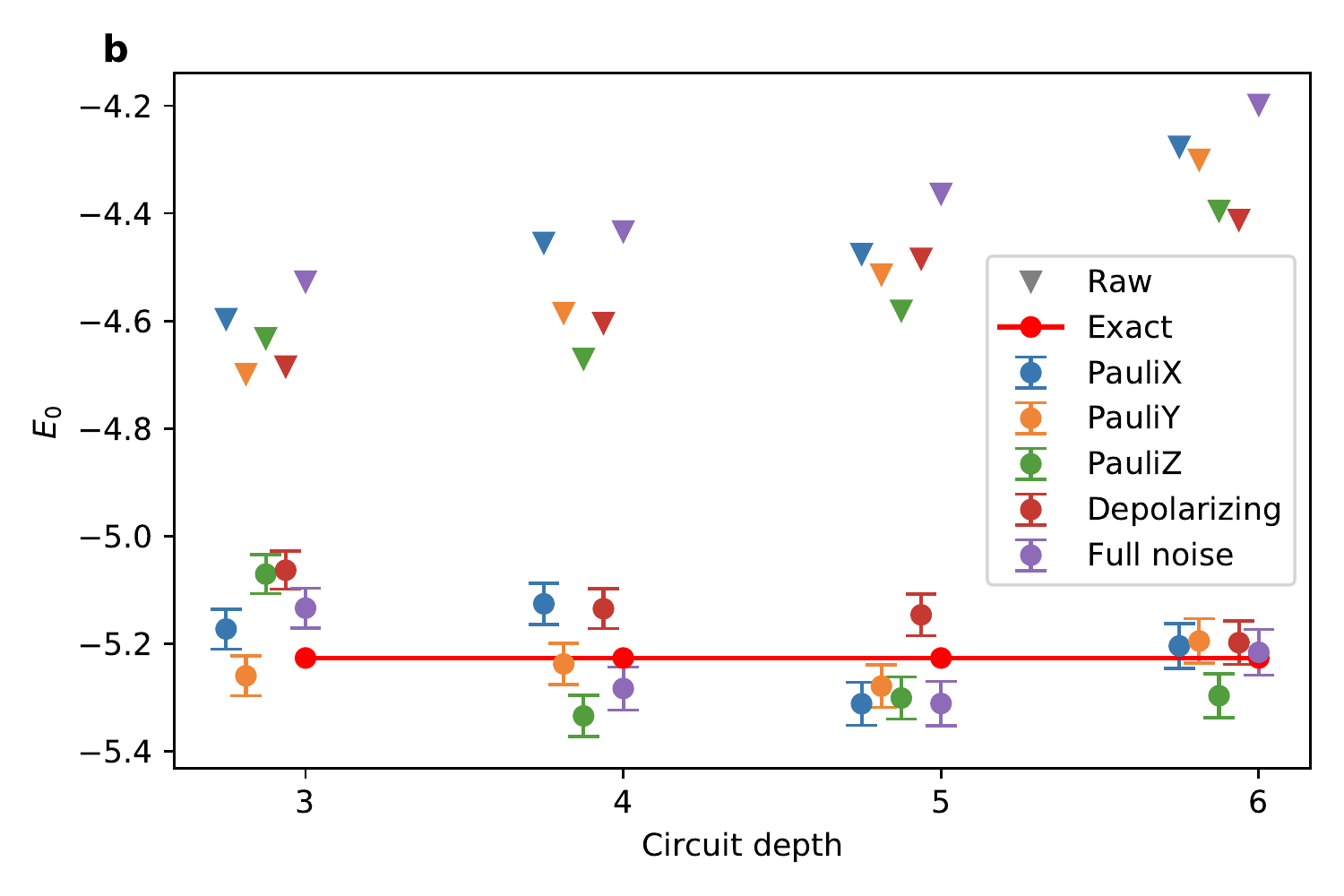}
  \caption{Numerical results for the ground-state energy $E_0$ as a function of the coupling strength $x$ of the Hamiltonian in Eq.~\eqref{eq:transversal_ising_hamiltonian} (left panel, \textbf{a}) and as a function of the circuit depth of the VQE ansatz for $x=-1$ (right panel, \textbf{b}). The number of qubits is (\textbf{a}) $N=2$ and (\textbf{b}) $N=4$ respectively. We plot results for five different kinds of quantum noise, including the Pauli channels with $\Lambda = X$ (blue points), $\Lambda = Y$ (orange points), $\Lambda = Z$ (green points) after a single-qubit gate, the depolarizing channel (brown points) after a CNOT gate, and the full noise model (purple points) simulated for the \textit{ibmq$\_$santiago} device. The triangles (points with error bar) represent  the raw (mitigated) data. For better visibility, data points for the same parameter set are offset horizontally. The red solid line corresponds to the results obtained from exact diagonalization.}
  \label{fig:numerical-mitigation}
\end{figure}

Figure~\ref{fig:numerical-mitigation} shows results obtained from classically simulating a noisy quantum device and applying our mitigation technique for different values of the coupling strength (left panel) and different circuit depths (right panel). The purities $\tr(\rho'^2)$ were evaluated using quantum state tomography~\cite{Nielsen2000}. We see that after the mitigation, our results for the variational energy converge to the true ground-state energy extracted from exact diagonalization within statistical errors. In Fig.~\ref{fig:numerical-mitigation}a, we observe that for decreasing values of the coupling $x$, the mitigated results converge better to the true ground-state energy. This is because the accuracy $\epsilon$ is proportional to the norm $||H||_{\infty}$ of the Hamiltonian $H$, see Eq.~\eqref{eq:L}, and  $||H||_{\infty}$ becomes smaller for smaller $x$. In Fig.~\ref{fig:numerical-mitigation}b, we see that the variational energy converges better to the true ground-state energy when using deeper circuits. This is because the square of the accuracy $\epsilon$ is proportional to $L^{-1}$, see Eq.~\eqref{eq:L}.

\begin{figure}[htp!]
\centering
  \includegraphics[width = 0.48\columnwidth]{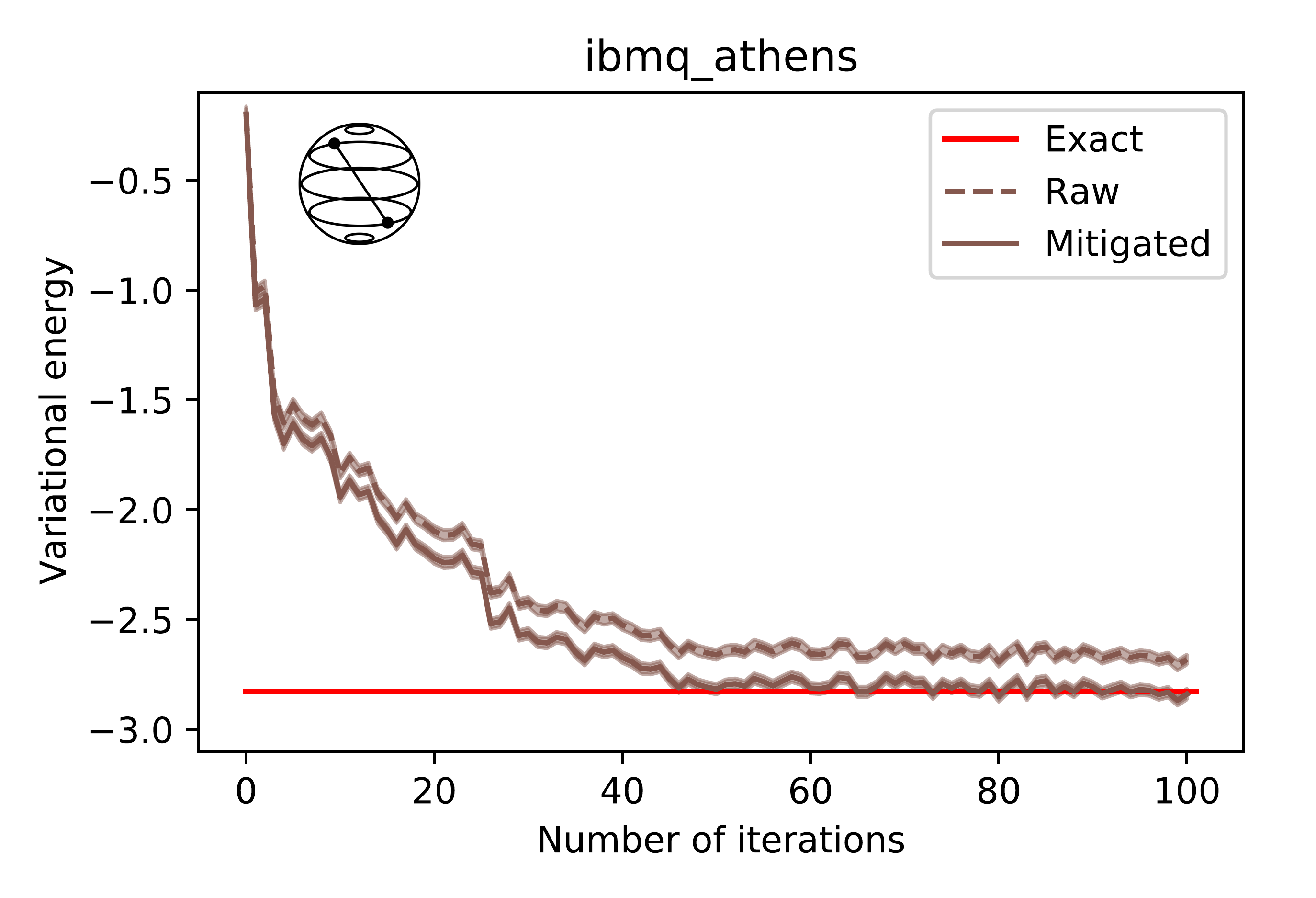}
  \includegraphics[width = 0.48\columnwidth]{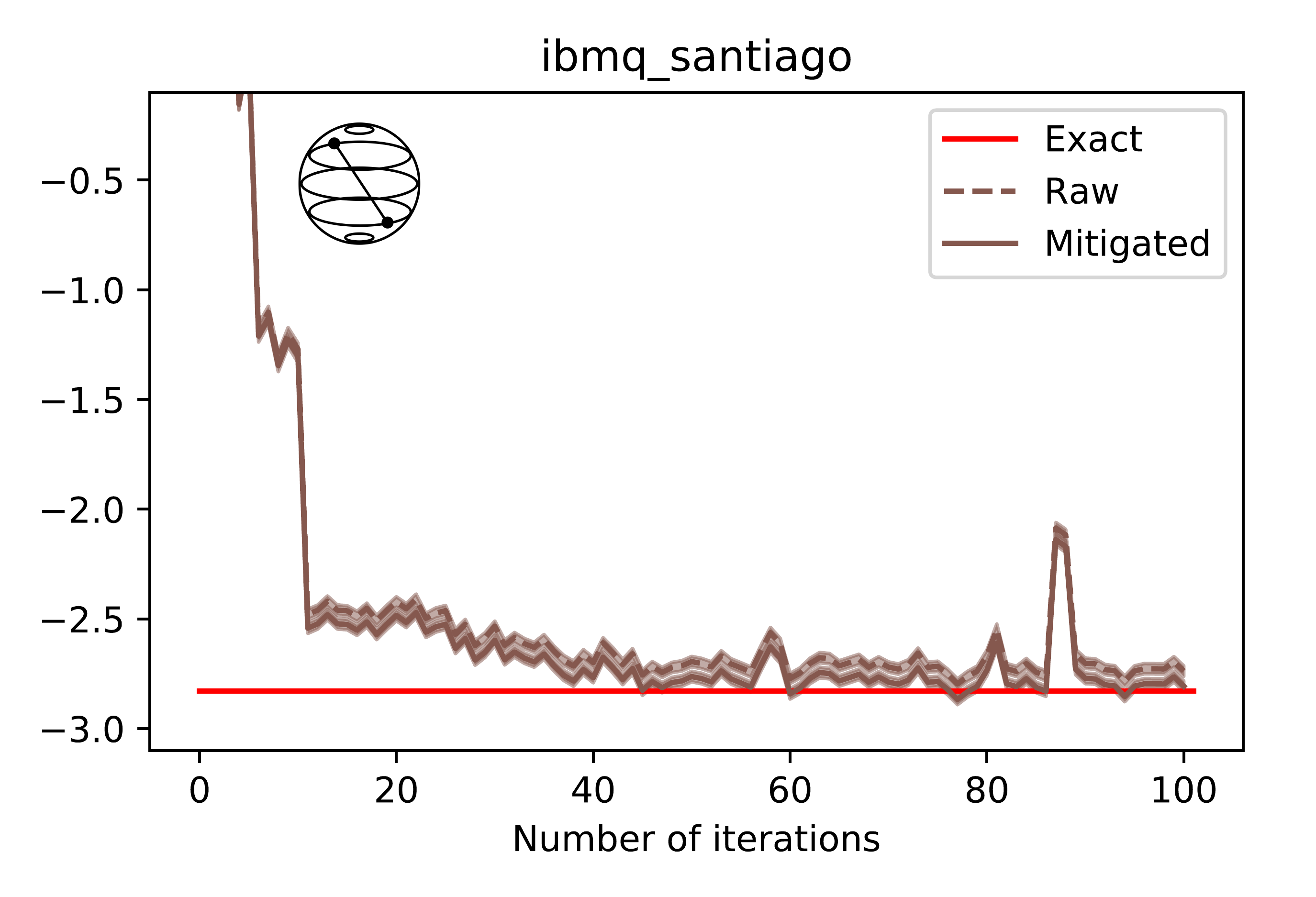}
  \caption{Experimental results for the variational energy $E(\mbox{\boldmath$\theta$})$ of the Hamiltonian in Eq.~\eqref{eq:transversal_ising_hamiltonian} with $x=-1$, obtained with running the VQE algorithm with 100 iterations on the IBM quantum devices \textit{ibmq\_athens} (left panel) and \textit{ibmq\_santiago} (right panel). We plot the raw experimental data (dashed brown line), the mitigated data (solid brown line), and the results obtained from exact diagonalization (solid red line). The mitigated data is obtained by multiplying the raw data by a factor of $(1-r)^{-1}$, see Eq.~\eqref{eq:mitigation-formula}, where $r$ is derived by measuring the purity $\tr(\rho'^2)$ of the state $\rho'$ at the last step of the VQE iteration using quantum state tomography. The red solid line corresponds to the results obtained from exact diagonalization of Eq.~\eqref{eq:transversal_ising_hamiltonian}.}
  \label{fig:Q2device}
\end{figure}

We finally carry out experiments on real quantum hardware with $N=2$ qubits. In Fig.~\ref{fig:Q2device}, we plot the VQE descent curves using the \textit{ibmq\_athens} (left panel) and the \textit{ibmq\_santiago} (right panel) quantum devices. We see that our mitigation scheme works well even on real quantum hardware with both coherent and incoherent noise. This suggests that the levels of coherent noise on these quantum devices are relatively small compared to those of the incoherent noise. 

\section{Conclusion and outlook}

In this work, we presented a new method to mitigate incoherent noise in parametric quantum circuits, by projecting the quantum noise channel onto a depolarizing noise channel. This projection is based on the assumption that the layered parametric ansatz circuits, which are widely used in variational quantum algorithms, are sufficiently close to unitary 2-designs, such that the Pauli noise is projected onto a depolarizing channel. We corroborated this assumption by carrying out proof-of-principle numerical experiments, and we examined the scalability of the method. We also demonstrated the performance of the mitigation protocol on quantum hardware, focusing on small-scale problems, for which we observe rapid convergence to the exact results.

In future work, we will investigate which of the underlying assumptions of our mitigation scheme could be relaxed, in order to generalize the method. For example, the LTM noise assumption could be relaxed by implementing a perturbative expansion of $\Lambda$ in Eq.~\eqref{eq:pauli-channel}, as used in most RB protocols. Moreover, the requirement on the total error rate $pL\ll 1$ could also be relaxed by taking into account higher-order contributions in Eq.~\eqref{eq:1} and expressing them in a form similar to Eq.~\eqref{eq:U-2-design}. Finally, the exponential overhead of evaluating the purity $\tr(\rho'^2)$ could be circumvented by measuring the error rate $p$ in Eq.~\eqref{eq:final-noisy-result} using RB protocols.

\acknowledgments
X.W.\ and X.F.\ are supported in part by the NSFC of China under Grant No.\ 11775002, No.\ 12070131001, and No.\ 12125501, and by the National Key Research and Development Program of China under Contract No.\ 2020YFA0406400. Research at Perimeter Institute is supported in part by the Government of Canada through the Department of Innovation, Science and Industry Canada and by the Province of Ontario through the Ministry of Colleges and Universities.
L.F.\ is partially supported by the U.S.\ Department of Energy, Office of Science, National Quantum Information Science Research Centers, Co-design Center for Quantum Advantage (C$^2$QA) under contract number DE-SC0012704, by the DOE QuantiSED Consortium under subcontract number 675352, by the National Science Foundation under Cooperative Agreement PHY-2019786 (The NSF AI Institute for Artificial Intelligence and Fundamental Interactions, http://iaifi.org/), and by the U.S.\ Department of Energy, Office of Science, Office of Nuclear Physics under grant contract numbers DE-SC0011090 and DE-SC0021006.
S.K.\ acknowledges financial support from the Cyprus Research and Innovation Foundation under project ``Future-proofing Scientific Applications for the Supercomputers of Tomorrow (FAST)'', contract no.\ COMPLEMENTARY/0916/0048.
G.P.\ is financially supported by the Cyprus  Research and Innovation Foundation under contract number POST-DOC/0718/0100 and from project NextQCD, co-funded by the European Regional Development Fund and the Republic of Cyprus through the Research and Innovation Foundation with contract id EXCELLENCE/0918/0129.  We acknowledge the use of IBM Quantum services for this work. The views expressed are those of the authors, and do not reflect the official policy or position of IBM or the IBM Quantum team. P.S.\ acknowledges support from Agencia Estatal de Investigaciín (the R\&D project CEX2019-000910-S, funded by MCIN/ AEI/10.13039/501100011033, Plan National FIDEUA PID2019-106901GB-I00, FPI), Fundació Privada Cellex, Fundació Mir-Puig, and from Generalitat de Catalunya (AGAUR Grant No. 2017 SGR 1341, CERCA program).

\bibliographystyle{JHEP}
\bibliography{Papers.bib}

\end{document}